\documentclass[apj,numberedappendix]{emulateapj}

\usepackage{epsfig}
\usepackage{amsmath}
\usepackage{enumerate}
\usepackage{mathrsfs}

\usepackage{natbib}
\usepackage{hyperref}

\usepackage{ulem}
\usepackage[squaren, Gray, cdot]{SIunits}
\usepackage{color}
\newbox\grsign \setbox\grsign=\hbox{$>$} \newdimen\grdimen \grdimen=\ht\grsign
\newbox\simlessbox \newbox\simgreatbox \newbox\simpropbox
\setbox\simgreatbox=\hbox{\raise.5ex\hbox{$>$}\llap
     {\lower.5ex\hbox{$\sim$}}}\ht1=\grdimen\dp1=0pt
\setbox\simlessbox=\hbox{\raise.5ex\hbox{$<$}\llap
     {\lower.5ex\hbox{$\sim$}}}\ht2=\grdimen\dp2=0pt
\setbox\simpropbox=\hbox{\raise.5ex\hbox{$\propto$}\llap
     {\lower.5ex\hbox{$\sim$}}}\ht2=\grdimen\dp2=0pt
\def\simgt{\mathrel{\copy\simgreatbox}}
\def\simlt{\mathrel{\copy\simlessbox}}

\def\T{{\mathcal T}}
\def\scr{s_{\rm cr}}
\def\Ts{T_{\rm s}}
\def\Tm{T_{\rm melt}}
\def\zmelt{z_{\rm melt}}
\def\beq{\begin{equation}}
\def\eeq{\end{equation}}
\def\sel{s_{\rm el}}
\def\spl{s_{\rm pl}}
\def\dspl{\dot{s}_{\rm pl}}
\def\sigcr{\sigma_{\rm cr}}
\def\Um{U_{\rm melt}}
\def\Uth{U_{\rm th}}
\def\Eq{Equation}
\def\Eqs{Equations}
\def\Ekin{E_{\rm kin}}
\def\Emag{E_{\rm mag}}
\def\Eel{E_{\rm el}}
\def\zd{z_{\rm damp}}
\def\Eaft{E_{\rm aft}}
\def\tc{t_{\rm cond}}
\def\LL{L}
\def\lw{l}
\def\Lum{\mathscr L}

\newcommand{\md}{\mathrm{d}}
\newcommand{\alfven}{Alfv\'{e}n }

\begin{document}

\title{Plastic damping  of Alfv\'en waves in magnetar flares\\
and delayed afterglow emission}

\author{Xinyu Li and Andrei M. Beloborodov}
\affil{Physics Department and Columbia Astrophysics Laboratory, Columbia University, 
538 West 120th Street, New York, NY 10027}

\begin{abstract}
Magnetar flares generate \alfven waves bouncing in the closed magnetosphere with 
energy up to $\sim 10^{46}$~erg. We show that on a 10-ms timescale the waves are 
transmitted into the star and form a compressed packet of high energy density. This 
packet strongly shears the stellar crust and initiates a plastic flow, heating the crust 
and melting it hundreds of meters below the surface. A fraction of the deposited plastic 
heat is eventually conducted to the stellar surface, contributing to the surface afterglow 
months to years after the flare. A large fraction of heat is lost to neutrino emission or 
conducted into the core of the neutron star.
\medskip
\end{abstract}


%##############################################################

\section{Introduction}

Magnetars are luminous slowly rotating neutron stars powered by the decay of 
ulstrastrong magnetic fields $B=10^{14}-10^{16}\rm G$ (see e.g. 
\citet{2006csxs.book..547W}; \citet{2008A&ARv..15..225M} for reviews).
They have hot surfaces, produce nonthermal magnetospheric radiation
and strong bursts of hard X-rays.
Occasionally, magnetars produce giant flares with energies of $10^{44}-10^{46}\,\rm ergs$. 
To date three giant flares have been observed from three magnetars.
The main peak of the giant flare lasts $\sim 0.3$~s and can reach huge luminosities 
$\Lum\sim 10^{47}$~erg~s$^{-1}$.
Less powerful flares with $\Lum<10^{43}$~erg~s$^{-1}$ 
(often called ``bursts'') occur much more frequently.

The flares are associated with a sudden change in the magnetospheric configuration, 
which could be triggered by an instability inside or outside the neutron star 
\citep{1996ApJ...473..322T}. This cataclysmic event involves strong deviations from the 
magnetostatic equilibrium, launching waves of large amplitudes. Part of the released 
magnetic energy is promptly dissipated and converted to radiation, and part is stored 
in the excited waves. 

In particular, \alfven waves are generated with a total energy up to $\sim 10^{46}$~erg. 
They are trapped on the closed magnetic field lines, as the group velocity of \alfven 
waves is parallel to the magnetic field. The fate of their energy is poorly known. 
It was proposed that the \alfven waves can be damped through nonlinear processes 
\citep{1998PhRvD..57.3219T}, which become efficient at very large amplitudes of the 
waves (see Section~5.2).

In this paper, we propose another  mechanism of the \alfven wave dissipation, which 
results from the wave interaction with the star. The waves are ducted along the magnetic 
field lines with nearly speed of light and reach the stellar surface on a millisecond 
timescale. In Section~2 we examine the wave interaction with the star and find that a 
significant fraction of the wave energy is transmitted into the stellar crust. The reflected 
waves keep bouncing in the magnetosphere, however in a few tens of milliseconds most 
of their energy is drained and deposited into the crust, in the form of a compressed shear 
wave packet. In Section~3, we show that this packet causes strong plastic heating of the 
crust. In Section~4 we investigate the fate of heat deposited by the plastic damping of 
\alfven waves. In particular, we evaluate the heat flux conducted back to the surface and 
the resulting surface luminosity, which should emerge long after the flare. Our results are 
discussed in Section~5.

%#################################################################

\section{Wave transmission into the crust}
\label{dyn}

The crust is nearly incompressible and supports shear waves which can be excited 
by the \alfven waves impinging from the magnetosphere. Excitation of two-fluid
crustal modes can be neglected, and the recent claim that magnetospheric \alfven 
waves  transform into crustal Hall waves \citep{2015MNRAS.447.1407L} is incorrect.
Hall waves propagating parallel to the magnetic field ${\mathbf B}$ with frequency 
$\omega$ in the crust of density $\rho$ have refraction index 
${\cal N}=ck/\omega=\omega_{\mathrm{pe}}/\sqrt{\omega\omega_B}
\approx 10^{9}\, \rho_{11}^{1/2} \omega_5^{-1/2} B_{14}^{-1/2}$ 
(we use the standard notation $X_m=X/10^m$ for a quantity $X$ in cgs units). This 
implies a huge impedance mismatch with the magnetospheric \alfven waves, which 
have ${\cal N}\approx 1$, and therefore their transformation to Hall waves is 
suppressed.\footnote{Only electrons move in a Hall wave (analogous to whistler in 
     plasma physics) while ions are static. The velocity of the electron fluid 
     $\mathbf{v}_H=\mathbf{j}/en_e$ is related to electric current 
     $\mathbf{j}=(c/4\pi)\nabla\times {\mathbf B}$, which gives a tiny $v_{\rm H}$ 
     because of the high electron density $n_e$ in the crust. Therefore, the 
     ``two-fluid"(electron-ion) description is useful only for slow phenomena in the crust.}
No significant separation between electron and ion velocities can occur on ms timescales, 
and the response of the crust to the external disturbance is essentially single-fluid.

\subsection{Transmission coefficient}

Consider a magnetospheric \alfven wave of frequency $\omega$ impinging on the crust 
of the neutron star. For simplicity let us assume that the initial (unperturbed) magnetic 
field $B_z$ is uniform and vertical, so the wave is propagating vertically along the 
$z$-axis, and the horizontal displacement $\xi(z)$ is along the $y$-axis. 
The plasma-filled magnetosphere and the crust are excellent conductors; 
therefore the magnetic field is frozen in the medium and the horizontal field $B_y$ is 
related to the displacement by $B_y/B_z=\partial\xi/\partial z$.
The wave speed in the magnetosphere is close to the speed of light $c$, and the 
wavelength is $\lambda_0=2\pi c/\omega$. 

The wave propagation is described by the equation,
\begin{equation}
\label{eq:wave}
   \left[\rho(z)+\frac{B_z^2}{4\pi c^2}\right]\frac{\partial^2\xi}{\partial t^2} 
   = \frac{B_z^2}{4\pi}\frac{\partial^2\xi}{\partial z^2} - \frac{\partial\sigma}{\partial z}. 
\end{equation}
Here $\rho(z)$ is the mass density and $\rho+B_z^2/4\pi c^2$ can be thought of as the 
effective inertial mass density of the magnetized medium.\footnote{This expression is 
             approximate as it neglects the contribution from the horizontal field component 
             $B_y$. In the models presented below, $B_y>B_z$ when the wave propagates 
             into the crust; however, this only occurs in the dense region where 
             $B^2/4\pi\ll \rho c^2$ and the magnetic field inertia anyway may be neglected. 
             In the region where $B^2/4\pi\gg \rho c^2$ the wave amplitude $s=B_y/B_z<1$ 
             and it is acceptable to approximate $B^2\approx B_z^2$.}
The first term on the right-hand-side describes the restoring force of magnetic tension, 
and the last term describes the force due to the shear stress in the medium.

\begin{figure}
% \plotone{fig1.eps}
\vspace*{-0.3cm}
\begin{tabular}{c}
\includegraphics[width=0.48\textwidth]{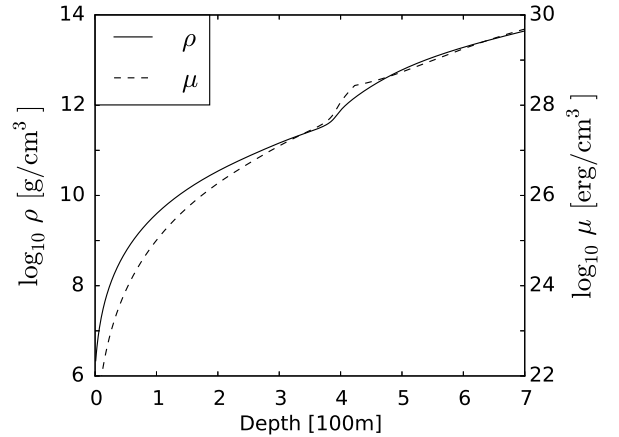} 
\end{tabular}
\caption{Density $\rho(z)$ and shear modulus $\mu(z)$ of the neutron star crust. 
The star is assumed to have mass $M=1.4 M_{\odot}$ and SLy equation of state.}
\label{fig1}
\end{figure}

In particular, if the medium is elastic with a shear modulus $\mu$ then $\sigma = -\mu s$, 
where $s=\partial\xi/\partial z$ is the strain of the elastic deformation. In this case, 
Equation~(\ref{eq:wave}) becomes a simple wave equation with the wave speed given by
\begin{equation}
  v^2(z)=\frac{B^2_z/4\pi+\mu(z)}{B^2_z/4\pi c^2+\rho(z)}. 
\end{equation}
In the magnetosphere, we will neglect the mass density $\rho$ and the shear modulus 
$\mu$, which gives $v=c$. In the crust, we will use the profiles $\rho(z)$ and $\mu(z)$ 
shown in Figure~\ref{fig1}. The density profile is obtained from the relativistic hydrostatic 
equation using SLy equation of state \citep{2004A&A...428..191H} for a neutron star 
with mass $M=1.4M_{\odot}$.\footnote{The
           ultrastrong magnetic field significantly changes pressure where the electron Fermi 
           energy is below the Landau energy. This impacts the density profile $\rho(z)$ at 
           shallow depths. However, at depths of interest in this paper (where 
           $\rho\gg 10^8$~g~cm$^{-3}$) this effect is small and neglected.}     
The radius of the star is $R=11.7\rm ~km$, and its surface gravitational acceleration is 
$g = (GM/R^2)(1-r_g/R)^{-1/2}=1.7\times 10^{14}$\, cm~s$^{-2}$ where $r_g=2GM/c^2$.
For the shear modulus $\mu$ we use the fitting formula given by 
\citet{2005ApJ...634L.153P} and \citet{2007MNRAS.375..261S} for low and high densities. 

As the wave propagates into the deeper crust, its speed is reduced and its wavelength 
is compressed,
\begin{equation}
   \lambda(z) = \lambda_0 \frac{v(z)}{c},  \qquad \lambda_0=\frac{2\pi c}{\omega}.
\end{equation}

\begin{figure}
% \plotone{fig2.eps}
\begin{tabular}{c}
\includegraphics[width=0.48\textwidth]{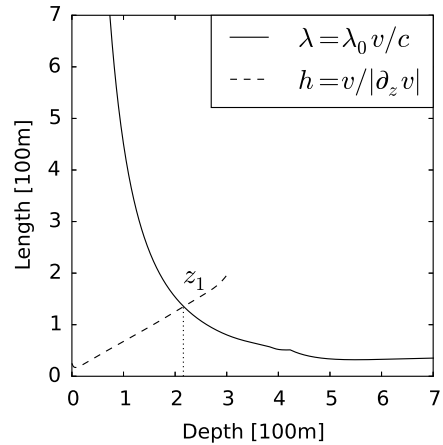} 
\end{tabular}
\caption{Solid curve shows the wavelength of the shear wave propagating in the 
magnetized crust, $\lambda = \lambda_0 v/c$, where $\lambda_0=10\,\rm km$ and 
$v$ is the speed of the wave. Dashed curve shows the characteristic scale-height of 
the wave deceleration, $h= v/|\partial_z v|$. A vertical magnetic field 
$B_z = 3\times 10^{14}\,\rm G$ is assumed in this example.}
\label{fig2}
\end{figure}

The reflection of the wave occurs in the region where the characteristic scale-height for 
the change of $v(z)$,
\begin{equation}
   h(z)=\frac{v}{|dv/dz|},
\end{equation}
is smaller than the wavelength $\lambda(z)$. Figure~\ref{fig2} shows $\lambda(z)$, 
$h(z)$, and the depth $z_1$ where they are equal. The typical value of $z_1$ is 
around 200 meters below the surface; its exact value depends on $B_z$.

%%%%%%%%%%%%%%%%%%%%%%%%%%
%%%%%%%%%%%%%%%%%%%%%%%%%%

%%%%%%%%%%%%%%%%%%%%%%
%%%%%%%%%%%%%%%%%%%%%%

The transmitted wave below $z_1$ has $\lambda\ll h$ and can be described in the 
WKB approximation. Then the wave displacement takes the form (e.g. \citet{f13}),
\begin{equation}
\label{eq:xi}
  \xi(z) = \frac{\rm const}{Z^{1/2}(z)}\cos\left[\omega\left( t-\int_0 ^z \frac{\md z'}{v(z')}\right)\right],
\end{equation}
where $Z(z)$ is the impedance,
\begin{equation}
\label{eq:Z}
  Z(z)=\left[\frac{B_z ^2}{4\pi}+\mu(z)\right]^{1/2}
          \left[\frac{B_z^2}{4\pi c^2}+\rho(z)\right]^{1/2}.
\end{equation}

\begin{figure}
% \plotone{fig3.eps}
\vspace*{-0.3cm}
\begin{tabular}{c}
\includegraphics[width=0.48\textwidth]{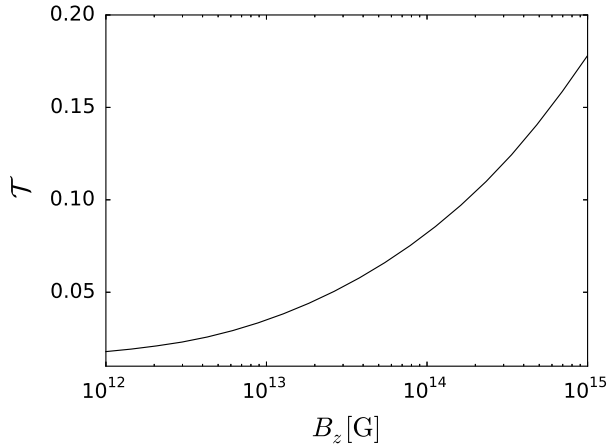} 
\end{tabular}
\caption{Transmission coefficient $\mathcal{T}$ as a function of vertical magnetic 
field $B_z$, obtained from the numerical solutions of Equation~(\ref{eq:wave}).}
\label{fig3}
\end{figure}

A simple estimate for the transmission coefficient is obtained using the impedance 
at $z_1$ \citep{1989ApJ...343..839B},
\begin{equation}
\label{eq:TC}
   \mathcal{T}\sim \frac{4Z(z_1)Z(0)}{\left[Z(z_1)+Z(0)\right]^2} 
   \approx \frac{4v(z_1)}{c}. 
\end{equation}
For instance, for $B_z=3\times 10^{14}\,\rm G$, Equation~(\ref{eq:TC}) gives 
$\mathcal{T}\sim 5\%$. A more accurate transmission coefficient is obtained by 
solving numerically the wave equation, which gives a higher value of $\T=12$\%
(the smoothness of the crustal density variation as the wave approaches $z_1$ 
enhances the transmission). The numerically calculated $\T(B_z)$ is shown in 
Figure~\ref{fig3}. It is comparable to 0.1 for typical magnetar fields.

%%%%%%%%%%% FIGURE %%%%%%%%%%%%%%%%%%

%%%%%%%%%%%%%%%%%%%%%%
%%%%%%%%%%%%%%%%%%%%%%

The reflection coefficient ${\mathcal R}=1-\T\sim 0.9$ is large, and the reflected Alfv\'en 
waves will bounce many times in the magnetosphere. Their amplitudes decrease by 
$\T\sim 10$\% every time they bounce from the surface. The repeated transmission 
events form a train of compressed waves in the crust. This train propagates into the 
crust with velocity $v\sim 10^{-2}c$.

One can show from Equation~(\ref{eq:xi}) that the strain $s=\partial\xi/\partial z$ in the 
transmitted wave evolves as $|s|\propto v^{-1}Z^{-1/2}\propto\rho^{1/4}$. It {\it increases} 
as the wave propagates into the deeper and denser crust. This has a simple physical 
reason: the wave decelerates, and hence its energy density $U_w$ grows as $v^{-1}$ 
(so that the wave continues to carry its energy flux $F_w=U_w v=const$). The wave 
energy oscillates between the kinetic energy and the horizontal field plus elastic energy 
of the crust. Therefore, $U_w$ may be written in two ways: 
$U_w\sim (\rho+B^2/4\pi c^2)\, \xi^2 \omega^2$ (kinetic) or 
$U_w\sim s^2 (B_z^2/4\pi+\mu)$ (magnetic+elastic). 
In the region where $\rho c^2>B^2/4\pi$ and $\mu<B_z^2/4\pi$ this requires 
$\xi^2\propto (\rho v)^{-1}$ and $s^2\propto v^{-1}$. Then the relation 
$s\sim \xi/\lambda\propto \xi/v$ gives 
\beq
 \xi\propto v^{1/2}, \qquad v\propto \rho^{-1/2}, \qquad  s\propto \rho^{1/4}.
\eeq 
In the lower crust where $\mu\simgt B_z^2/4\pi$ and 
$v\approx(\mu/\rho)^{1/2}\approx const \approx 10^8$~cm~s$^{-1}$ one finds 
$s\propto\rho^{-1/2}$. In this region the wave strain significantly decreases. 
This evolution of $s$ with depth (increase and then decrease) may be observed in
the numerical simulation presented below.

\subsection{Numerical model}

To illustrate the transmission process we set up a simple one-dimensional 
simulation of waves
bouncing in the magnetosphere between the footprints of a closed magnetic flux tube.
The \alfven waves are ducted along the magnetic field lines and 
the problem can be made one-dimensional by pretending that the flux tube is straight, 
and by placing its two opposite footprints on the $z$-axis, separated by distance $\LL$.
Here $\LL$ represents the length of magnetospheric field lines.
The stellar crust with the density profile $\rho(z)$ is placed symmetrically at the two ends 
of the computational box. The crust thickness $\sim 1$~km is much smaller than $\LL$.

For any initial shear distortion of the field lines, one can calculate the subsequent 
dynamics of the generated \alfven waves by solving numerically Equation~(\ref{eq:wave}). 
In our numerical models, we take the initial distortion of the form, 
\begin{equation}
    \xi_0(z) = A \exp\left[-\frac{(z-z_0)^2}{2\lw^2}\right],
\end{equation}
which is localized in the middle of the box $z_0$, far away from the crust, with 
$\lw< \LL$.
The distortion immediately splits into two waves propagating toward the opposite ends of 
the box. The strain profile of each wave $s(z)=\partial\xi/\partial z$ is determined by the 
initial distortion. An important parameter of the wave is its initial maximum strain,
\beq
   s_0=\frac{1}{2}\,\max |\partial_z \xi_0(z)|=\frac{A}{2\lw}\,e^{-1/2}.
\eeq
The total energy initially stored in the two waves is
\begin{equation}
    E_{0}=\frac{\sqrt{\pi}}{8\pi}\,B_z^2\,\frac{A^2S}{\lw},
\end{equation}
where $S$ is the cross section area of the flux tube.

%%%%%%%%%%%%%%%%%%%%%%
\begin{figure}
% \plotone{fig4.eps}
\vspace*{-0.3cm}
\begin{tabular}{c}
\includegraphics[width=0.48\textwidth]{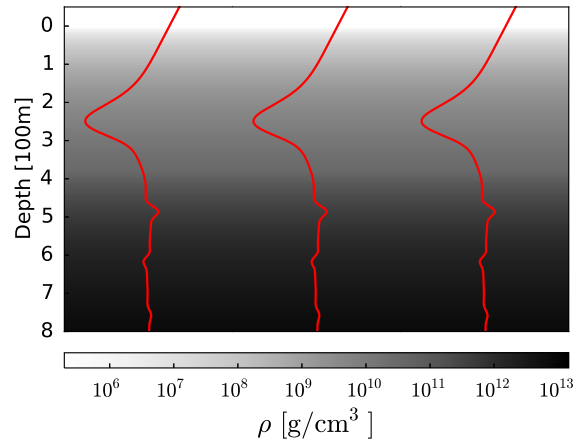} 
\end{tabular}
\caption{A snapshot of the wave after four reflection/transmission events. 
Solid red curves show the magnetic field lines deformed by the horizontal 
displacements in the wave; background grey color shows the density of the crust. 
The wave front has reached the depth of 800~m by this time. One can see the four 
oscillations in the transmitted and compressed packet. The shape of each oscillation 
reflects the initial shape of the magnetospheric wave assumed in the simulation.
This simulation included no plastic dissipation and assumed the magnetic field 
$B_z = 3\times 10^{14}\,\rm G$.}
\label{fig4}
\end{figure}
%%%%%%%%%%%%%%%%%%%%%%

We follow the evolution of the waves and their interaction with the crust until almost all 
wave energy $E_0$ has been drained from the magnetosphere; this typically takes tens 
of light-crossing times $\LL/c$. The wave equation~(\ref{eq:wave}) is solved on a grid 
with 1000 points in the magnetosphere (uniformly spaced) and a much finer grid in the 
crust (one point per meter). Convergence tests have been done to ensure that the grid 
is sufficiently large to resolve the wave dynamics.

The snapshot of the simulation in Figure~\ref{fig4} shows the distortion of the crust 
at time $t=4.6\LL/c$, when the magnetospheric waves have bounced four times. 
The parameters of this sample model are $\LL=40$~km, $\lw=5/\sqrt{2}$~km, 
$A=5$~km, and $B_z=3\times 10^{14}$~G. The corresponding 
$s_0=(2e)^{-1/2}\approx 0.43$ and $E_0/S\approx 4.5\times 10^{32}$~erg~cm$^{-2}$.
In the snapshot shown in the figure, about 1/3 of the wave energy $E_0$ has already 
been transmitted into the crust. The transmitted wave has been decelerated to 
$v\approx 10^8$~cm~s$^{-1}$ and compressed by the factor of 
$c/v\approx 3\times 10^2$. 
The compression creates a high energy density of the horizontal magnetic field 
at $z=200-500$~m, $(sB_z)^2/8\pi\sim 10 (s_0B_z)^2/8\pi$, and strain $s\sim 3s_0$.

\medskip

%###########################################################

\section{Plastic Heating}

The description of the wave dynamics in Section~\ref{dyn} is incomplete because it 
assumes the elastic response $\sigma=-\mu s$ everywhere in the crust.
The more realistic model must take into account two facts:
(1) When the solid crust is deformed by the shear wave beyond a critical stress 
$\sigcr$ its response becomes plastic rather than elastic.
(2) The crustal temperature may be high enough to reduce $\sigcr$ or even melt the 
crust, leading to $\sigcr\approx 0$.
Therefore, the model should keep track of the crustal temperature.

\subsection{Pre-flare temperature profile}

The typical persistent surface temperature of magnetars is $\Ts\sim (3-4)\times 10^6$~K
\citep{2006csxs.book..547W}. It corresponds to the radiation energy flux 
$F=\sigma \Ts^4\sim 10^{22}$~erg~s$^{-1}$, where 
$\sigma=5.67\times 10^{-5}$~erg~s$^{-1}$~cm$^{-2}$~K$^{-4}$ is the Stefan-Boltzmann 
constant. A usual way to estimate the subsurface temperature profile of neutron 
stars $T(z)$ assumes that the surface flux $F$ is supplied by quasi-steady diffusion 
of heat from the crust. Then $T(z)$ is given by the equation,
\begin{equation}
\label{eq:Tin}
   \kappa(T,z) \frac{dT}{dz} =F=\sigma \Ts^4,
\end{equation}
where $\kappa$ is the effective conductivity, which is dominated by degenerate electrons 
at densities $\rho>10^6$~g~cm$^{-3}$ and by radiation in the low density layers near the 
surface. Note that $\kappa$ depends on the local magnetic field. 
Combining \Eq~(\ref{eq:Tin}) with the hydrostatic equation $dP/dz=\rho g$ gives
\begin{equation}
\label{eq:Tstruct}
   \frac{\md \log T}{\md \log P}=\frac{3}{16}\frac{PK}{g}\frac{\Ts^4}{T^4}.
\end{equation}
Here $P$ is the pressure, 
$g=(GM/R^2)(1-r_g/R)^{-1/2}$
is the surface gravitational acceleration,
and $K=16\sigma T^3 /3\kappa \rho$ is the effective opacity; all quantities are 
measured in the local rest frame of the crust. The surface luminosity and temperature 
measured by a distant observer are $\Lum^\infty = (1-r_g/R)\Lum$ and 
$T_s^{\infty} = (1-r_g/R)^{1/2} T_s$ \citep{1977ApJ...212..825T}.

Equation~(\ref{eq:Tstruct}) assumes that the temperature profile had enough time to relax 
to the steady state at depths of interest, which typically takes $\sim 1$~yr. 
In a true steady state, the relatively high surface temperature of magnetars requires 
a source of heat at depths of a few hundred meters \citep{2006MNRAS.371..477K}. 
Alternatively, one may view this temperature profile as qausi-steady,
slowly cooling after a previous heating episode.

If one accepts this thermal model for the pre-flare state of the crust, one can find $T(z)$ 
from Equation~(\ref{eq:Tstruct}) and determine the melting depth $\zmelt$ above which 
the crust is melted. We use the code of \citet{1999A&A...351..787P} to calculate the 
thermal conductivity and the melting point $\Tm(\rho)$ of the crustal material. 
An approximate result is sufficient for the purposes of this paper and we do not 
discuss here the poorly known chemical composition of the magnetar crust. 
For simplicity, we assume an iron crust with a small impurity parameter.

We also assume that the pre-flare magnetic field is not far from vertical. This is a 
reasonable assumption for the melted layer ($z\simlt 100$~m, see below) where the 
crustal magnetic field should match the force-free magnetosphere. 
A strong toroidal field can only be stored in the core of the neutron star or the deep crust.

Equation~(\ref{eq:Tstruct}) can be solved numerically as described in detail in previous 
works, which calculated the relation between the surface effective temperature $T_s$ 
and the internal temperature $T_b$ measured at neutron-drip depth 
$z_b\approx 400$~m ($\rho_b = 4\times10^{11}\,\rm g$~cm$^{-3}$). In particular, 
\citet{2001A&A...374..213P} provide a fitting formula for $T_b(T_s)$ for various magnetic 
fields and assuming an iron crust. We use this relation to impose the condition $T=T_b$ 
at $z=z_b$, and then reconstruct the profile $T(z)$ in the region of interest 
$\rho>10^8$~g~cm$^{-3}$ (above or below $z_b$) by integrating \Eq~(\ref{eq:Tstruct}) 
from $z_b$. Thus we avoid integration in the shallow surface layers where thermal 
conductivity is dominated by radiation, and so we only use the electron conductivity in 
our calculations. 

For a given $\Ts$, this calculation gives the subsurface temperature profile $T(z)$ and 
the melting depth $\zmelt$. The melting temperature is approximately given by 
\beq
  \Tm\approx 2.4\times 10^9\rho_{12}^{1/3} {\rm ~K}.
\eeq 
The exact value of melting depth $\zmelt$ depends on the magnetic field and its 
orientation relative to the stellar surface. A strong field increases the thermal conductivity 
along ${\mathbf B}$ and decreases it perpendicular to ${\mathbf B}$. Therefore, a 
vertical field tends to reduce the internal temperature, thus decreasing $\zmelt$. 
A horizontal field would hamper the heat flow in the vertical direction and increase 
$\zmelt$. A typical $\zmelt$ in magnetars with non-horizontal surface fields is 
$\sim 100$~m.

\subsection{Plastic flow}

As the wave packet propagates into the crust below $\zmelt$, it starts to interact with the 
solid phase (lattice). The response of the lattice is elastic as long as its strain is below a 
critical value $\scr$. The maximum $\scr\sim 0.1$ is comparable to the yielding threshold 
for an ideal crystal \citep{2010MNRAS.407L..54C}. The actual strain in the wave 
$s\sim s_0(\T c/v)^{1/2}$ is much higher than $\scr$, and so 
the wave initiates a strong plastic flow with the high frequency $\omega$.
In contrast to fluid \alfven wave or elastic shear wave, the plastic flow is dissipative, i.e. it 
converts the wave energy to heat, reducing its amplitude.
Below we include this process in our wave propagation model.

The plastic heating rate per unit volume is 
\beq 
    \frac{d\Uth}{dt}=-\sigma\dspl, 
\eeq
where $\spl=s-\sel$ is the plastic part of the strain, $\sel$ is the elastic part, and 
$\sigma$ is the shear stress sustained by the plastic flow.
In the plastic regime $|\sigma|>\sigcr$ where $\sigcr=\mu\scr$. 
The simple model of ``viscoplastic solid'' (e.g. \citet{ir08}) gives the stress of the plastic 
flow in the form,
\beq
   |\sigma|=\sigcr+\eta|\dspl|, 
\eeq
where $\eta$ is a viscosity coefficient.

The crystal becomes ``soft'' (i.e. $\sigcr$ drops) if it is heated to a temperature 
comparable to the melting point $\Tm$. The softening effect is responsible for the 
thermoplastic instability that can release internal magnetic stresses in magnetars 
\citep{2014ApJ...794L..24B}.
This instability however develops on a timescale much longer than 10~ms and 
does not affect the dynamics considered in this paper. Here the plastic flow is driven 
by the strong external magnetic stress from the flare (rather than develops 
spontaneously inside the crust) and immediately reaches huge strains 
$|s|\gg\scr$ and high temperatures.
Because $|s|\gg \scr$, the detailed behavior of $\scr(T)$ and $\sigcr(T)$ is not important; 
our calculation should merely take into account the fact that plastic heating switches off 
when $T$ approaches $\Tm$.

This effect is included as follows: the stress $\sigma$ of the plastic flow is multiplied by 
the factor $1-U_{\rm th}/U_{\rm melt}$, where $\Uth(\rho,T)$ is the thermal energy 
density and $\Um=\Uth(\rho,\Tm)$. This prescription enforces $\sigma=0$ when 
$T=\Tm$.

The stress in the elastic regime $\sigma=-\mu s$ must match the plastic stress at 
$s=\scr$. This condition is automatically satisfied for the cold crystal. For a hot crystal the 
reduction of $\sigcr(T)$ may be interpreted as the reduction of shear modulus $\mu$ or 
the reduction of $\scr$ (or both). These details are not important for our model, because 
the plastic flow has $|s|\gg \scr$. The numerical models presented below assume 
$\scr(T)=0.1=const$ and use the following prescription for $\sigma$,
\begin{equation}
\label{eq:sigma}
\sigma =
  \left(1-\frac{\Uth}{\Um}\right)  \times
   \left\{
  \begin{array}{ll}
     -\mu s, &  \textrm{elastic}\\
   \left(0.1
    \mu +\eta \dspl\right) {\rm sign}(-s),
    & \textrm{plastic}\\
    0, & \textrm{liquid}
  \end{array}
\right.
\end{equation}
where $\mu$ is the shear modulus at $T\ll\Tm$ shown in Figure~1, and one may think of 
$(1-\Uth/\Um)\mu$ as the shear modulus reduced by heating. We verified that practically 
the same results are obtained if we choose a temperature-dependent 
$\scr=0.1(1-\Uth/\Um)$ with shear modulus unchanged by heating.

Finally, we must choose $\eta$, which is unknown for the crustal material.
The transition between the plastic and elastic regimes is smooth if $\eta$ vanishes 
when $|s|=\scr$. Therefore, we assume $\eta$ of the form $\eta=\alpha\,\mu\,|\,|s|-\scr|$, 
where $\alpha$ is a constant. We tried various values of $\alpha$ and found that plastic 
heating weakly depends on it as long as $\alpha$ is sufficiently large, 
$\alpha>3\times 10^{-6}$~s. Our sample numerical models use 
$\alpha=3\times 10^{-5}$~s.

The dynamic system described by \Eq~(\ref{eq:wave}) with $\sigma$ given by 
\Eq~(\ref{eq:sigma}) satisfies the energy conservation law,
\beq
\label{eq:energy}
   \frac{dQ}{dt}= S \int \frac{d\Uth}{dt}\, \md z =-\frac{d}{dt}\left(\Ekin+E_B+\Eel\right),
\eeq
where
\begin{eqnarray}
  \Ekin   &=& S \int \left(\rho+\frac{B_z^2}{4\pi c^2}\right) \frac{\dot{\xi}^2}{2}\, \md z,\\
  E_B &=& S \int \frac{s^2 B_z^2}{8\pi}\, \md z, \\
  \Eel    &=& S \int \frac{\mu\,\sel^2}{2}\, \md z,
\end{eqnarray}
where $|\sel|<\scr$ in the elastic zone and $|\sel|=\scr$ in the plastic zone.

The plastic flow occurs where $|\sigma|$ exceeds $\sigcr$ and continues as long as 
$d|s|/dt>0$. Whenever the local absolute value of the strain stops growing, the plastic 
flow switches to the elastic regime; at this point $\sel$ and $\sigma$ are reset to zero.

%%%%%%%%%%%%%%%%%%
\begin{figure}
% \plotone{fig5.eps}
\vspace*{-0.3cm}
\begin{tabular}{c}
\includegraphics[width=0.48\textwidth]{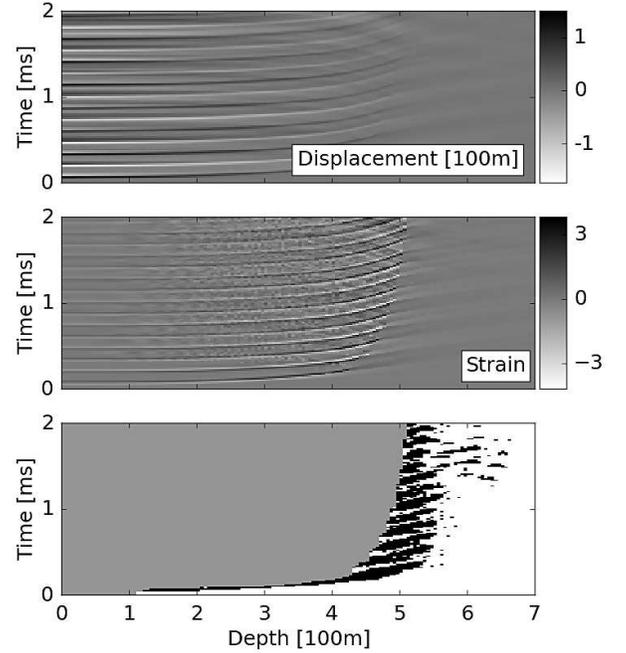} 
\end{tabular}
\caption{Shear wave propagation in the magnetar crust viewed on the spacetime 
diagram: Upper panel: horizontal displacement of the wave, $\xi$. 
Middle panel: strain $s=\partial\xi/\partial z$.
Lower panel shows where the crust is deformed elastically (white), flowing plastically 
(black), and melted (gray).}
\label{fig5}
\end{figure}
%%%%%%%%%%%%%%%%%%

\subsection{Wave damping and post-flare crustal temperature}

We re-run the model described in Section~2.2 with the new expression for $\sigma$ 
that takes into account the plastic damping in the crust (\Eq~(\ref{eq:sigma})).
The initial state is assumed to have the surface temperature $T_s=3\times 10^6$~K.
All other parameters are the same as in Section~2.2, in particular 
$B_z=3\times 10^{14}$~G and $s_0=(2e)^{-1/2}\approx 0.43$.
The spacetime diagram of the wave evolution in the crust is presented in Figure~\ref{fig5}.
It shows the wave displacement and strain, and indicates the elastic, plastic, and melted 
regions.

\begin{figure}
% \plotone{fig6.eps}
\vspace*{-0.3cm}
\begin{tabular}{c}
\includegraphics[width=0.48\textwidth]{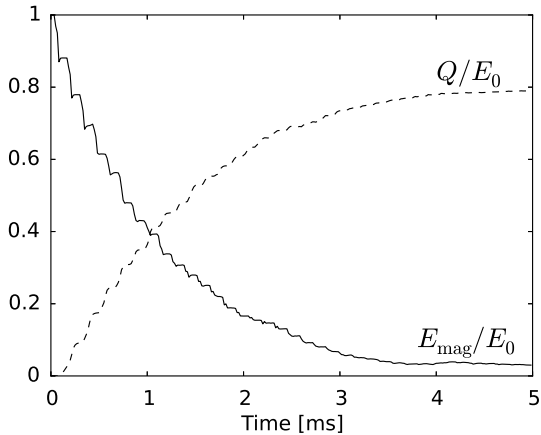} 
\end{tabular}
\caption{Solid curve shows the evolution of the energy fraction left in the magnetosphere 
and dashed curve shows the energy fraction converted to heat. 
Each step in the solid curve corresponds to the simultaneous reflection of the two 
symmetric waves bouncing in the magnetosphere in the opposite directions. 
The reflection coefficient ${\cal R}\approx 0.88$, and the magnetosphere loses 
energy as ${\cal R}^{N}$ where $N$ is the number of reflection events.
Each step takes $\LL/c=(4/30)$~ms, so $\Emag/E_0\approx 0.88^{t/0.133\rm ms}$.}
\label{fig6}
\end{figure}

Figure~\ref{fig6} shows the history of the wave energy transmission from the 
magnetosphere to the crust and the plastic damping effect. One can see that most of 
the transmitted wave energy is promptly converted to heat. We have verified that our 
numerical simulation satisfies the conservation law (\Eq~(\ref{eq:energy})) with 
accuracy better than 1\%. Each time the wave hits the surface, the transmission 
coefficient is approximately 12\%, and almost all the wave energy $E_0$ is damped 
after $\sim 10$~ms, so most of $E_0$ becomes stored as crustal heat. This heating 
results in deep melting of the crust, down to 500~m.

%%%%%%%%%%%%%%%%%%
%%%%%%%%%%%%%%%%%%

%%%%%%%%%%%%%%%%%%%%%%%%
\begin{figure} 
% \plotone{fig7_1.eps}
% \plotone{fig7_2.eps}
\vspace*{-0.3cm}
\begin{tabular}{c}
\includegraphics[width=0.48\textwidth]{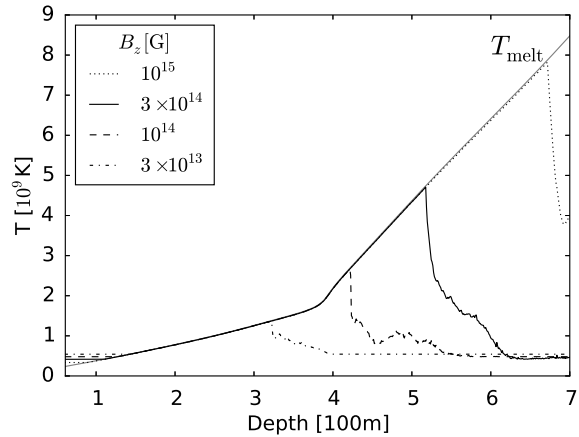} \\
\includegraphics[width=0.48\textwidth]{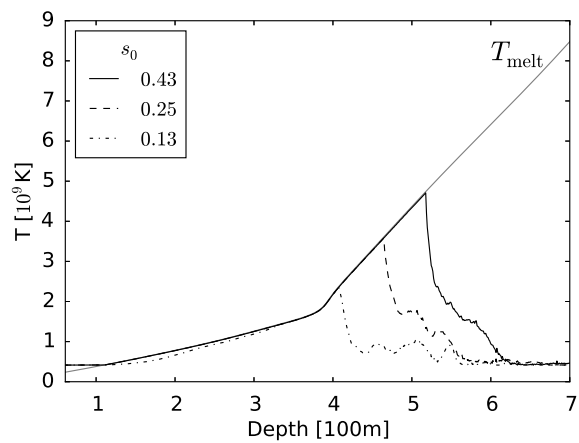} 
\end{tabular}
\caption{Temperature profile after the magnetospheric \alfven waves have been 
absorbed by the crust; upper panel for fixed $s_0=0.43$ and different $B_z$, and 
lower panel for fixed  $B_z=3\times 10^{14}$~G and different $s_0$. 
The left boundary in the figure is chosen at depth
$z\approx 60$~m where $\rho = 10^9$~g~cm$^{-3}$;
at $z\simlt 100$~m the crust is melted before the flare,
and hence no plastic heating can take place.}
\label{fig7}
\end{figure}
%%%%%%%%%%%%%%%%%%%%%%%%

Since plastic dissipation switches off at the meting point, the crust naturally acquires the 
``ceiling'' temperature $T\approx T_m$ in an extended region below the surface.
The resulting temperature profile immediately after the flare is shown in Figure~\ref{fig7}.
To investigate how the results depend on $B_z$ and $s_0$, we have calculated the 
models with $B_z/10^{14}{\rm ~G}=0.3,1,3,10$ and $s_0=0.13,0.25,0.43$. Stronger 
waves in stronger magnetic fields melt deeper layers of the crust, up to 600~m in 
the calculated models.

\medskip

%####################################################

\section{Cooling}

After the flare, the hot crust will cool on a much longer timescale. 
Two main processes cool the crust: neutrino emission and heat conduction.
The temperature evolution with time is described by the following equation,
\begin{equation}
\label{eq:cool}
C_V \frac{\partial T}{\partial t} = 
  \frac{\partial}{\partial z}\left(\kappa
  \frac{\partial T}{\partial z}\right)-
  \dot{q}_\nu,
\end{equation}
where $\kappa$ is the thermal conductivity and $C_V$ is the heat capacity of the
crust; both are functions of local $\rho(z)$, $T(z,t)$, and ${\mathbf B}$. 
The sample numerical models presented below assume a uniform vertical magnetic 
field $B=B_z=const$. We use $\kappa(\rho,T,B)$ and $C_V(\rho,T,B)$ calculated by 
the code of \citet{1999A&A...351..787P}. The term $\dot{q}_\nu(\rho,T,B)$ is the rate 
of local cooling by neutrino emission. This rate is described in detail by 
\citet{2001A&A...374..213P}. They provide useful analytical approximations for 
four relevant channels of neutrino emission: plasmon decay, bremsstrahlung, 
synchrotron, and electron-positron annihilation. We use their formulas in our calculations.

Similar to previous simulations of time-dependent heat diffusion in magnetars 
\citep{2006MNRAS.371..477K,2009ApJ...698.1020B,2009A&A...496..207P},
we separate the crust into two regions: a blanketing envelope and 
an interior region. Here we choose the envelope boundary at $z_b\approx 60$~m where 
$\rho=\rho_b=10^{9}\rm ~g/cm^3$.
The typical timescale of heat diffusion from this depth is $t_b \ll 10^6 s$.
It is sufficiently short to give a quasi-steady state in the envelope, and so the steady-state 
solution may be used to determine the relation between $T_b=T(z_b)$ and the effective 
surface temperature $T_s$.
Note that $T_s$ defines the energy flux $F=\sigma T_s^4$ through the envelope $z<z_b$, 
and thus in essence the $T_b$-$T_s$ relation is a relation between $T_b$ and the heat 
flux $F=\kappa\, \partial T/\partial z$ at $z_b$. It serves as a boundary condition for our 
time-dependent heat diffusion problem at $z>z_b$. Since this boundary condition relies 
on the steady-state solution at $z<z_b$ it can only be accurate when $T(t,z>z_b)$
evolves on timescales longer than $t_b$.

We calculated the $T_b$-$T_s$ relation at $\rho_b=10^9$~g~cm$^{-3}$ using the 
steady-state solutions obtained in Section~3.1. This gave a tabulated boundary 
condition $F(T_b)$ at the upper boundary of our computational box $z_b\approx 60$~m. 
The lower boundary is chosen at $z\approx 1$~km, near the bottom of the crust where 
$\rho\sim 10^{14}$~g~cm$^{-3}$. The exact position of the lower boundary is not important 
as long as it is deep enough. The deep crust has a high thermal conductivity and the heat 
is absorbed by the (approximately isothermal) core of a huge heat capacity. We use the 
absorbing boundary condition at constant temperature $T\sim 3\times 10^8$~K, 
neglecting the increase of the core temperature due to the absorbed heat.

The initial condition $T(z,0)$ for \Eq~(\ref{eq:cool}) is provided by the plastic heating model 
(Figure~\ref{fig7}). The initial temperature increases along the melting curve $T_m(z)$, 
reaches maximum, and drops at larger depths. 
We evolve this initial temperature profile on a uniform grid of 1000 points and a time step 
of 10~s for $10^7$ steps. To speed up the simulation, the values of $\kappa$, $C_V$, and 
$\dot{q}_\nu$ on the grid are updated every 500 time steps.
Convergence tests verified that this resolution is sufficient to obtain accurate results.
We also verified that our simulation conserves energy with better than 1\% accuracy.
The thermal energy lost by the crust is partially carried away by neutrinos and partially 
conducted through the boundaries.  

\begin{figure}
% \plotone{fig8.eps}
\vspace*{-0.3cm}
\begin{tabular}{c}
\includegraphics[width=0.48\textwidth]{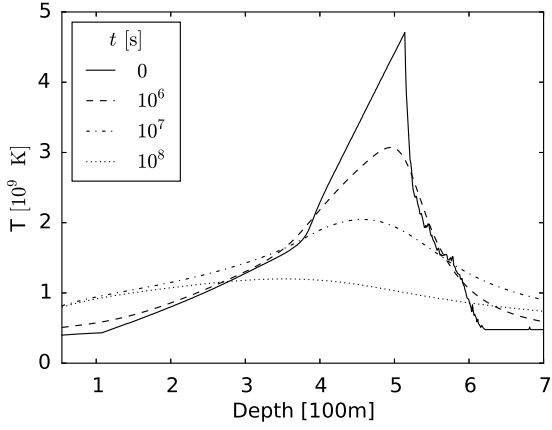} 
\end{tabular}
\caption{Evolution of the crustal temperature profiles in our fiducial flare model 
with $B=B_z=3\times 10^{14}$~G and $s_0=0.43$.}
\label{fig8}
\end{figure}

Figure~\ref{fig8} shows the gradual evolution of the temperature profile $T(z)$ after 
the flare with the fiducial parameters (see Section~2.2 and Figures~5, 6). 
During the first month, the initial peak of temperature at $z\sim 500$~m is reduced from 
$\sim 5\times 10^9$~K mainly due to neutrino losses. Then the peak continues to flatten 
and spread due to thermal conduction, forming a rather flat profile of $T\simlt 10^9$~K 
in a few years.

%%%%%%%%%%%%%%%%%%%%%%%%
%%%%%%%%%%%%%%%%%%%%%%%%

We find that plasmon decay and bremsstrahlung make the dominant contributions to 
neutrino cooling, and synchrotron neutrino emission becomes significant in stronger 
magnetic fields $B_z\sim 10^{15}$~G. Electron-positron annihilation dominates neutrino 
cooling only in the shallow, low-density region of the crust, and its net contribution to the 
energy loss is negligible.

%%%%%%%%%%%%%%%%%%%%%%%%
\begin{figure}
% \plotone{fig9.eps}
\vspace*{-0.3cm}
\begin{tabular}{c}
\includegraphics[width=0.48\textwidth]{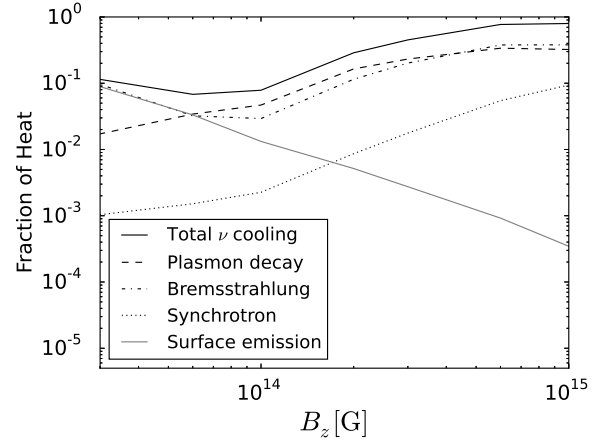} 
\end{tabular}
\caption{Fraction of the post-flare crustal heat lost through surface emission and various 
channels of neutrino emission, as a function of $B=B_z$. The flare is assumed to excite a 
pair of \alfven waves with $s_0=0.43$ (Section~2.2) which are plastically damped in the crust.}
\label{fig9}
\end{figure}
%%%%%%%%%%%%%%%%%%%%%%%%

%%%%%%%%%%%%%%%%%%%%%%%%
\begin{figure}
% \plotone{fig10_1.eps}
% \plotone{fig10_2.eps}
\vspace*{-0.3cm}
\begin{tabular}{c}
\includegraphics[width=0.48\textwidth]{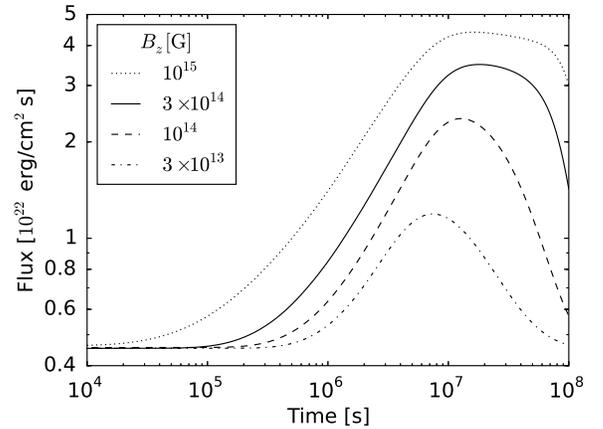} \\
\includegraphics[width=0.48\textwidth]{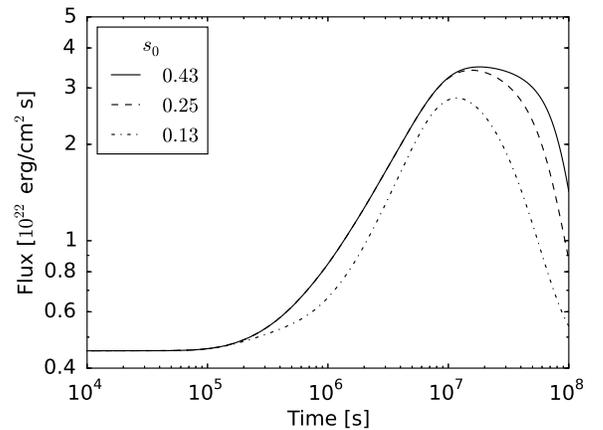} 
\end{tabular}
\caption{Surface thermal flux caused by the plastic heating in the giant flare. 
Upper panel: fixed $s_0=0.43$ and varying $B=B_z$.
Lower panel: fixed $B_z=3\times 10^{14}$~G and varying $s_0$.}
\label{fig10}
\end{figure}
%%%%%%%%%%%%%%%%%%%%%%%%

With increasing $B_z$ (and at fixed amplitude $s_0$ of the waves excited in the flare) 
the deposited plastic heat increases, which increases the role of neutrino cooling.
As a result, the relative contributions of plasmon decay, bremsstrahlung, and synchrotron 
neutrino emission depend on $B_z$. This dependence is shown in Figure~\ref{fig9}, 
where we also show the energy fraction that is conducted to the stellar surface and 
radiated away. The remaining energy fraction (not shown in Figure~9) is conducted into 
the core of the neutron star.

One can see from Figure~\ref{fig9} that only a small fraction of the stored crustal 
heat is conducted to and radiated from the stellar surface.
For example, at $B_z=3\times 10^{14}\,\rm G$ less than $1\%$ is conducted to the 
surface; roughly half of heat is lost neutrino emission and half is conducted to the core.

The heat radiated from the surface produces a delayed afterglow emission of the 
flare. The solution of \Eq~(\ref{eq:cool}) gives the surface radiation flux $F$ as a function 
of time. This flux is shown in Figure~\ref{fig10}. One can see that the surface flux peaks 
with a significant delay after the flare --- it takes the thermal conduction timescale 
(of months to years) to transport the crustal heat to the surface.

The core remains much colder than the plastically heated crust. The heat conducted to the 
core cannot significantly boost its temperature because (1) the core has a large heat 
capacity, so a huge energy $E_{\rm th}\sim 10^{48}T_{9}^2$~erg would be required to
heat it to e.g. $10^9$~K \citep{2004ARA&A..42..169Y}, and (2) the core is efficiently 
cooled by neutrino emission, and the cooling rate quickly grows at high temperatures.

%#####################################################
\bigskip

\section{Discussion}

\subsection{Plastic damping and cooling}

In this paper we described the phenomenon of plastic damping of \alfven waves 
generated in magnetar flares. Our results may be summarized as follows.
\medskip

(1) {\it Transmission.}
The flare generates magnetospheric \alfven waves with energy density 
\beq
  U_0\sim \frac{\mu_B s_0^2}{2},
\eeq
where $\mu_B=B_z^2/4\pi$ is the tension of magnetic field lines, and $s_0\simgt 0.1$
is the shear strain of the field lines. The waves are quickly transmitted into the crust of 
the neutron star. The transmission coefficient is $\T\sim 0.1$ (Figure~3), and most of 
the wave energy is transmitted after $N\sim \T^{-1} \sim 10$ reflection events (Figure~6).
The transmitted waves form a train of $N$ oscillations propagating with velocity 
$v\simlt 10^{-2}c$ and compressed by the factor of $c/v$ (Figure~4).
\medskip

(2) {\it Compression.}
The wave energy, which is initially spread in the magnetosphere, upon transmission 
becomes compressed. The energy density of the transmitted wave is
\beq 
   U_w\approx \T\, \frac{c}{v} \, U_0, 
\eeq
where $v$ decreases to $10^8$~cm~s$^{-1}$ as the wave train propagates toward the 
bottom of the crust. The transmission occurs at depths $z$ of a few hundred meters 
where the crustal density $\rho\sim 10^{10}-10^{11}$~g~cm$^{-3}$. In this region, $U_w$ 
exceeds the maximum energy that could be stored in the elastic deformation of the crust, 
$U_{\rm el}=\mu \scr^2/2$, and the wave propagation is still sustained by the tension of 
magnetic field lines, $\mu_B$. Therefore, the transmission also leads to the strain 
amplification: $s^2/s_0^2\approx U_w/U_0\sim 10$.
\medskip

(3) {\it Plastic flow.}
The shear strain of the transmitted wave, $s\sim (\T c/v)^{1/2} s_0$,
exceeds the maximum possible strain of elastic deformation $\scr\sim 0.1$. 
Therefore, the wave induces a strong plastic flow of the crust, which
dissipates the wave energy. A minimum plastic stress $\sigma$ is 
comparable to $\mu\scr$ and may be higher, as the shearing motion is very fast and the 
viscoplastic term $\eta\dot{s}$ can dominate the stress. 
As the wave propagates into denser layers $\rho\sim 10^{12}$~g~cm$^{-3}$, 
the shear modulus increases to $\mu\approx 10^{28}\rho_{12}$~erg~cm$^{-3}$ (Figure~1).
The plastic heat density deposited by the wave train is given by
\beq
   \Uth \sim \sigma s N \simgt \mu \scr s N.
\eeq  
The high $\Uth$ given by this estimate implies that the wave energy density $U_w$ 
converts to $\Uth$, i.e. efficient damping occurs.
\medskip

(4) {\it Melting.}
Damping of the wave is buffered by melting --- plastic damping is inefficient where the 
heated crust becomes nearly liquid, and the wave continues to propagate to denser 
layers that have a higher $\Tm(\rho)$. As a result, a simple temperature profile 
$T\approx \Tm(\rho)$ is created by the plastic flow in an extended region of the crust 
(Figure~7). 

Most of the wave damping occurs at depth $\zd$ where $\Tm$ is so high that the wave 
dissipation becomes marginally capable of melting the crust. Thus $\zd$ is also the 
depth of the melted region. At this depth the following condition is satisfied,
\beq
   C_V\Tm\sim U_w.
\eeq
We found $\zd\sim 500$~m for a typical wave energy in magnetar giant flares and 
$B_z\sim 3\times 10^{14}$~G (Figure~7). Deeper melting $\zd\sim 700$~m is possible 
if the giant flare occurs in a flux rope of a particularly strong field $B_z\simgt 10^{15}$~G.
\medskip

(5) {\it Cooling.}
On a timescale of months to years, the deposited heat is mostly lost to neutrino emission 
and conducted into the core of the star (Figures~8 and 9). A modest energy $\Eaft$ is 
conducted to the stellar surface and emitted in a delayed afterglow radiation. A typical 
energy radiated per unit area is $\Eaft/S\sim 10^{30}$~erg~cm$^{-2}$.
The timescale for the rise of afterglow luminosity is the thermal conduction time $\tc\sim 10^7$~s. In a broad range of the flare parameters, the peak flux of surface afterglow is $F_{\max}\sim (2-4)\times 10^{22}$~erg~cm$^{-2}$~s$^{-1}$ (Figure~10).

There are ways to refine our model of surface afterglow from plastic damping of 
magnetospheric waves. All sample models shown in this paper assumed approximately 
vertical (radial) magnetic field in the upper crust. A strongly inclined field would significantly 
reduce thermal conductivity in the radial direction and delay the crustal cooling. It could 
also bolster a high crustal temperature before the flare, which would give a deeper melted 
zone where plastic damping would be impossible. In this case, the flare could only cause 
heating of the deep crust where practically all heat is wasted to neutrino emission and 
inward conduction. Thus, a strong non-radial field component tends to reduce the 
expected afterglow emission.

Our presented models assumed iron composition everywhere in the crust, including 
the blanketing envelope. Light element composition of the envelope would increase its 
thermal conductivity \citep{2003ApJ...594..404P}, decreasing the internal temperature 
and reducing the depth of the melted layer in the pre-flare crust. Therefore, if magnetars 
have a light element envelope, their post-flare cooling occurs faster. This effect somewhat 
increases the afterglow flux, especially at early times, and may offset the opposite effect 
of the non-radial magnetic field.

\subsection{Other mechanisms of \alfven wave damping}

Nonlinear interactions of \alfven waves in the magnetosphere provide an additional damping mechanism. The existing estimates \citep{1998PhRvD..57.3219T} suggest that this mechanism will be dominant at very high amplitudes of the waves, $s_0\simgt 1$.
The nonlinear interactions occur as the \alfven waves bounce from the stellar surface and collide in the magnetosphere. The nonlinear terms in the electrodynamic equations show two types of wave interactions: 

\noindent
(1) $A+A\rightarrow F$: two \alfven waves $A$ convert into a fast magnetosonic 
wave $F$ (which may escape the magnetosphere). 
The damping of \alfven waves by this ``3-wave'' interaction occurs on the timescale,
\beq
\label{eq:3wave}
  t_{\rm damp}\sim \frac{\lambda/c}{(k_{\bot}\xi)^2}\sim 
   \left(\frac{k_\parallel}{k_\perp}\right)^2\,\frac{\lambda/c}{s^2},
\eeq
where $\lambda=2\pi/k_\parallel$ is the wavelength along ${\mathbf B}$ (comparable to 
the length of the magnetospheric field line $L$), $\xi$ is the characteristic displacement in 
the waves, and $k_\perp$ is the wavevector component perpendicular to the magnetic 
field. The \alfven waves, which are ducted along the curved magnetic field lines, may be 
expected to have $k_\perp\sim k_\parallel$.
  
 \noindent
(2)  $A+A\rightarrow A+A$:  two \alfven waves generate two new \alfven waves. This 
``4-wave'' interaction initiates a cascade to high $k_\perp$, which may lead to the wave 
dissipation on small scales \citep{1998PhRvD..57.3219T}.
The damping time due to this higher-order process is 
\beq
 \label{eq:4wave}
  t_{\rm damp}\sim \frac{\lambda/c}{(k_{\bot}\xi)^4}
     \sim  \left(\frac{k_\parallel}{k_\perp}\right)^4 \frac{\lambda/c}{s^4}.
\eeq
The time $t_{\rm damp}$ given by \Eqs~(\ref{eq:3wave}) and (\ref{eq:4wave}) should be 
compared with $\T^{-1}\LL/c\sim 10 \LL/c$, the lifetime of the \alfven waves to transmission 
and plastic damping in the crust. The numerical coefficients in \Eqs~(\ref{eq:3wave}) and 
(\ref{eq:4wave}) have not been calculated, however the estimates suggest that if the flare 
generates $s_0\simgt 1$, the nonlinear wave interactions can reduce $s_0$ to a value 
$\simlt 1$ before the waves are damped plastically in the crust.
 
The wave cannot be completely damped by the plastic mechanism. 
In particular, at strains $|s|<\scr$ it propagates with no significant damping. 
The residual wave train will reach the bottom of the crust and enter the liquid core. 
It will travel through the core along the magnetic field lines and after time $\sim 2r/v$ 
(typically shorter than 1~s) the train will again emerge somewhere at the bottom of the 
crust and continue to propagate upward.

The low-amplitude waves will continue to travel through the magnetosphere 
and the star for a while. Their lifetime at any given amplitude $s$ is limited by the nonlinear 
interactions in the magnetosphere $t_{\rm damp}\propto s^{-2}$. 
The \alfven waves are also subject to gradual ohmic dissipation, as their propagation 
involves excitation of electric currents demanded by $\nabla\times{\mathbf B}\neq 0$.
After the flare, the effective resistivity of the magnetosphere is controlled by the threshold 
voltage of electron-positron discharge that organizes to conduct the electric currents 
\citep{2007ApJ...657..967B}.

\subsection{Observed afterglow}

Sudden crustal heating followed by gradual crustal cooling was proposed to power the 
afterglow of the giant flare in SGR~1900+14 \citep{2002ApJ...580L..69L}.
The afterglow was extremely bright in the first hours after the flare, 
$\Lum\sim 10^{37}-10^{38}$~erg~s$^{-1}$, and during the next month it showed 
a power law decay $\Lum\propto t^{-0.7}$ \citep{2001ApJ...552..748W}.
\citet{2002ApJ...580L..69L} explored how heat should be deposited to give the observed 
afterglow light curve and found that heating should be approximately uniform throughout 
the 500-m-deep layer below the surface. This implies, in particular, enormous heating in 
the shallow layers $z\ll 100$~m. The heating mechanism in the low density layers is 
unclear and certainly cannot be provided by plastic dissipation. Therefore we do not 
attempt to explain the early afterglow of SGR~1900+14 by crustal heating. We also 
note that the afterglow spectrum was nonthermal \citep{2001ApJ...552..748W}, which 
suggests a magnetospheric source. 

Plastic damping of magnetospheric \alfven waves produces a well defined temperature 
profile  of the crust: $T\approx \Tm(z)$ down to $\zd$. This leads to specific predictions 
for the afterglow light curves (Figure~10), with the surface flux 
$F\sim (2-4)\times 10^{22}$~erg~cm~$^{-2}$~s$^{-1}$ on a timescale $\simgt 100$~d. 
This flux and timescale appear to be consistent with observations of some less 
energetic ``transient'' magnetars after their bursting activity.

In particular, the luminosity of SGR~1627-41 after its outbursts in 1998 and 2008 showed 
a decay on a year timescale 
\citep{2006A&A...450..759M,2008MNRAS.390L..34E,2012ApJ...757...68A}. The
luminosity at $t\sim 100$~d was 
$\Lum\sim 7\times 10^{34}(d/11{\rm ~kpc})^2$~erg~s$^{-1}$ after the 1998 outburst 
and $\Lum\sim 2\times 10^{34}(d/11{\rm ~kpc})^2$~erg~s$^{-1}$ after 
the 2008 outburst, where the distance $d\approx 11$~kpc was inferred from the apparent 
location of SGR~1627-41 in a star-forming region (Hurley et al. 1999). The decay on a 
year timescale is consistent with the crust melting down to $\zd\sim 300$~m, and the 
observed luminosity $\Lum$ is consistent with the melted crust area occupying 
$\sim 10$\% of the stellar surface.

Swift~J1822.3-1606 provides another example. It produced afterglow emission following 
the outburst in 2011 \citep{2012ApJ...754...27R,2012ApJ...761...66S,2014ApJ...786...62S}.
Similar to the afterglow of SGR~1627-41, its light curve may be described as a double 
exponential, with the second (longer) exponential component visible after $\sim 100$~d. 
\citet{2014ApJ...786...62S} used a crustal cooling model to describe both the early and 
late afterglow components in Swift~J1822.3-1606. In their model, heat deposition is a 
phenomenological parameter adjusted to reproduced observations. We find that
plastic damping of magnetospheric waves is only capable of explaining the late
afterglow component, and the early component must invoke a different heat source.
The late component has the luminosity and decay timescale similar to those observed in 
SGR~1627-41, consistent with the crust melting down to $\zd\sim 300$~m.

A reliable identification of the crustal afterglow is complicated by the presence of another, 
{\it nonthermal}, emission component. The nonthermal source is likely present during the 
afterglow of SGR~1627-41 \citep{2012ApJ...757...68A}, and nonthermal hard X-rays are 
unambiguously detected in the transient magnetar 1E~1547.0-5408 during its afterglow 
following the 2009 outburst \citep{2012MNRAS.427.2824E,2012ApJ...748..133K}.
The nonthermal activity is usually associated with the twisted equilibrium magnetosphere, 
which carries persistent electric currents \citep{2002ApJ...574..332T,2013ApJ...762...13B}. 
The twist is ohmically dissipated over a year timescale, which happens to be comparable 
to the timescale of crustal cooling.

Another complication is the expected {\it external} heating of the stellar surface bombarded 
by magnetospheric particles. This heating occurs at the footprint of the current-carrying 
magnetic field lines (``j-bundle''). As the magnetosphere slowly untwists, the j-bundle
shrinks and so does its hot footprint \citep{2009ApJ...703.1044B}. Such shrinking hot spots 
have been observed in several transient magnetars, including the canonical transient 
magnetar XTE~J1810-197. Following an outburst in 2003 it showed an X-ray afterglow 
decaying on a year timescale, with luminosity $\Lum\sim 2\times 10^{34}$~erg~s$^{-1}$ 
at $t\sim 1$~yr \citep{2007Ap&SS.308...79G}. The observed area $A(t)$ and luminosity 
$\Lum(t)$ of the hot spot evolved in agreement with the predictions of the untwisting 
magnetosphere model. Similar shrinking hot spots were observed in 1E~1547.0-5408, 
CXOU~J164710.2-455216, SGR~0501+4516, SGR~0418+5729 (see the data collection 
in \citet{2011heep.conf..299B} and references therein) and more recently in 
Swift~J1822.3-1606 \citep{2014ApJ...786...62S} and the Galactic Center magnetar 
SGR~J1745-2900 \citep{2015MNRAS.449.2685C}.

Strong \alfven waves and deep plastic heating are certainly expected in energetic events, 
in particular in giant flares. All three giant flares observed to date were emitted by 
persistently active magnetars, which maintain a high level of both magnetospheric activity 
and surface luminosity. It is possible that plastic damping of \alfven waves is the main 
mechanism that keeps the crust hot in these objects.

\acknowledgments
This work was supported by NASA grant NNX13AI34G.

\bibliography{ms}

\begin{thebibliography}{}
\expandafter\ifx\csname natexlab\endcsname\relax\def\natexlab#1{#1}\fi

\bibitem[{{An} {et~al.}(2012){An}, {Kaspi}, {Tomsick}, {Cumming}, {Bodaghee},
  {Gotthelf}, \& {Rahoui}}]{2012ApJ...757...68A}
{An}, H., {Kaspi}, V.~M., {Tomsick}, J.~A., {et~al.} 2012, \apj, 757, 68

\bibitem[{{Beloborodov}(2009)}]{2009ApJ...703.1044B}
{Beloborodov}, A.~M. 2009, \apj, 703, 1044

\bibitem[{{Beloborodov}(2011)}]{2011heep.conf..299B}
{Beloborodov}, A.~M. 2011, in High-Energy Emission from Pulsars and their
  Systems, ed. D.~F. {Torres} \& N.~{Rea}, 299

\bibitem[{{Beloborodov}(2013)}]{2013ApJ...762...13B}
---. 2013, \apj, 762, 13

\bibitem[{{Beloborodov} \& {Levin}(2014)}]{2014ApJ...794L..24B}
{Beloborodov}, A.~M., \& {Levin}, Y. 2014, \apjl, 794, L24

\bibitem[{{Beloborodov} \& {Thompson}(2007)}]{2007ApJ...657..967B}
{Beloborodov}, A.~M., \& {Thompson}, C. 2007, \apj, 657, 967

\bibitem[{{Blaes} {et~al.}(1989){Blaes}, {Blandford}, {Goldreich}, \&
  {Madau}}]{1989ApJ...343..839B}
{Blaes}, O., {Blandford}, R., {Goldreich}, P., \& {Madau}, P. 1989, \apj, 343,
  839

\bibitem[{{Brown} \& {Cumming}(2009)}]{2009ApJ...698.1020B}
{Brown}, E.~F., \& {Cumming}, A. 2009, \apj, 698, 1020

\bibitem[{{Chugunov} \& {Horowitz}(2010)}]{2010MNRAS.407L..54C}
{Chugunov}, A.~I., \& {Horowitz}, C.~J. 2010, \mnras, 407, L54

\bibitem[{{Coti Zelati} {et~al.}(2015){Coti Zelati}, {Rea}, {Papitto},
  {Vigan{\`o}}, {Pons}, {Turolla}, {Esposito}, {Haggard}, {Baganoff}, {Ponti},
  {Israel}, {Campana}, {Torres}, {Tiengo}, {Mereghetti}, {Perna}, {Zane},
  {Mignani}, {Possenti}, \& {Stella}}]{2015MNRAS.449.2685C}
{Coti Zelati}, F., {Rea}, N., {Papitto}, A., {et~al.} 2015, \mnras, 449, 2685

\bibitem[{{Enoto} {et~al.}(2012){Enoto}, {Nakagawa}, {Sakamoto}, \&
  {Makishima}}]{2012MNRAS.427.2824E}
{Enoto}, T., {Nakagawa}, Y.~E., {Sakamoto}, T., \& {Makishima}, K. 2012,
  \mnras, 427, 2824

\bibitem[{{Esposito} {et~al.}(2008){Esposito}, {Israel}, {Zane}, {Senziani},
  {Starling}, {Rea}, {Palmer}, {Gehrels}, {Tiengo}, {de Luca}, {G{\"o}tz},
  {Mereghetti}, {Romano}, {Sakamoto}, {Barthelmy}, {Stella}, {Turolla},
  {Feroci}, \& {Mangano}}]{2008MNRAS.390L..34E}
{Esposito}, P., {Israel}, G.~L., {Zane}, S., {et~al.} 2008, \mnras, 390, L34

\bibitem[{{Fitzpatrick}(2013)}]{f13}
{Fitzpatrick}, R. 2013, {Oscillations and Waves: An Introduction} (CRC Press)

\bibitem[{{Gotthelf} \& {Halpern}(2007)}]{2007Ap&SS.308...79G}
{Gotthelf}, E.~V., \& {Halpern}, J.~P. 2007, \apss, 308, 79

\bibitem[{{Haensel} \& {Potekhin}(2004)}]{2004A&A...428..191H}
{Haensel}, P., \& {Potekhin}, A.~Y. 2004, \aap, 428, 191

\bibitem[{{Irgens}(2008)}]{ir08}
{Irgens}, F. 2008, {Continuum Mechanics} (Springer)

\bibitem[{{Kaminker} {et~al.}(2006){Kaminker}, {Yakovlev}, {Potekhin},
  {Shibazaki}, {Shternin}, \& {Gnedin}}]{2006MNRAS.371..477K}
{Kaminker}, A.~D., {Yakovlev}, D.~G., {Potekhin}, A.~Y., {et~al.} 2006, \mnras,
  371, 477

\bibitem[{{Kuiper} {et~al.}(2012){Kuiper}, {Hermsen}, {den Hartog}, \&
  {Urama}}]{2012ApJ...748..133K}
{Kuiper}, L., {Hermsen}, W., {den Hartog}, P.~R., \& {Urama}, J.~O. 2012, \apj,
  748, 133

\bibitem[{{Lyubarsky} {et~al.}(2002){Lyubarsky}, {Eichler}, \&
  {Thompson}}]{2002ApJ...580L..69L}
{Lyubarsky}, Y., {Eichler}, D., \& {Thompson}, C. 2002, \apjl, 580, L69

\bibitem[{{Lyutikov}(2015)}]{2015MNRAS.447.1407L}
{Lyutikov}, M. 2015, \mnras, 447, 1407

\bibitem[{{Mereghetti}(2008)}]{2008A&ARv..15..225M}
{Mereghetti}, S. 2008, \aapr, 15, 225

\bibitem[{{Mereghetti} {et~al.}(2006){Mereghetti}, {Esposito}, {Tiengo},
  {Turolla}, {Zane}, {Stella}, {Israel}, {Feroci}, \&
  {Treves}}]{2006A&A...450..759M}
{Mereghetti}, S., {Esposito}, P., {Tiengo}, A., {et~al.} 2006, \aap, 450, 759

\bibitem[{{Piro}(2005)}]{2005ApJ...634L.153P}
{Piro}, A.~L. 2005, \apjl, 634, L153

\bibitem[{{Pons} {et~al.}(2009){Pons}, {Miralles}, \&
  {Geppert}}]{2009A&A...496..207P}
{Pons}, J.~A., {Miralles}, J.~A., \& {Geppert}, U. 2009, \aap, 496, 207

\bibitem[{{Potekhin}(1999)}]{1999A&A...351..787P}
{Potekhin}, A.~Y. 1999, \aap, 351, 787

\bibitem[{{Potekhin} \& {Yakovlev}(2001)}]{2001A&A...374..213P}
{Potekhin}, A.~Y., \& {Yakovlev}, D.~G. 2001, \aap, 374, 213

\bibitem[{{Potekhin} {et~al.}(2003){Potekhin}, {Yakovlev}, {Chabrier}, \&
  {Gnedin}}]{2003ApJ...594..404P}
{Potekhin}, A.~Y., {Yakovlev}, D.~G., {Chabrier}, G., \& {Gnedin}, O.~Y. 2003,
  \apj, 594, 404

\bibitem[{{Rea} {et~al.}(2012){Rea}, {Israel}, {Esposito}, {Pons},
  {Camero-Arranz}, {Mignani}, {Turolla}, {Zane}, {Burgay}, {Possenti},
  {Campana}, {Enoto}, {Gehrels}, {G{\"o}{\v g}{\"u}{\c s}}, {G{\"o}tz},
  {Kouveliotou}, {Makishima}, {Mereghetti}, {Oates}, {Palmer}, {Perna},
  {Stella}, \& {Tiengo}}]{2012ApJ...754...27R}
{Rea}, N., {Israel}, G.~L., {Esposito}, P., {et~al.} 2012, \apj, 754, 27

\bibitem[{{Scholz} {et~al.}(2014){Scholz}, {Kaspi}, \&
  {Cumming}}]{2014ApJ...786...62S}
{Scholz}, P., {Kaspi}, V.~M., \& {Cumming}, A. 2014, \apj, 786, 62

\bibitem[{{Scholz} {et~al.}(2012){Scholz}, {Ng}, {Livingstone}, {Kaspi},
  {Cumming}, \& {Archibald}}]{2012ApJ...761...66S}
{Scholz}, P., {Ng}, C.-Y., {Livingstone}, M.~A., {et~al.} 2012, \apj, 761, 66

\bibitem[{{Sotani} {et~al.}(2007){Sotani}, {Kokkotas}, \&
  {Stergioulas}}]{2007MNRAS.375..261S}
{Sotani}, H., {Kokkotas}, K.~D., \& {Stergioulas}, N. 2007, \mnras, 375, 261

\bibitem[{{Thompson} \& {Blaes}(1998)}]{1998PhRvD..57.3219T}
{Thompson}, C., \& {Blaes}, O. 1998, \prd, 57, 3219

\bibitem[{{Thompson} \& {Duncan}(1996)}]{1996ApJ...473..322T}
{Thompson}, C., \& {Duncan}, R.~C. 1996, \apj, 473, 322

\bibitem[{{Thompson} {et~al.}(2002){Thompson}, {Lyutikov}, \&
  {Kulkarni}}]{2002ApJ...574..332T}
{Thompson}, C., {Lyutikov}, M., \& {Kulkarni}, S.~R. 2002, \apj, 574, 332

\bibitem[{{Thorne}(1977)}]{1977ApJ...212..825T}
{Thorne}, K.~S. 1977, \apj, 212, 825

\bibitem[{{Woods} {et~al.}(2001){Woods}, {Kouveliotou}, {G{\"o}{\v g}{\"u}{\c
  s}}, {Finger}, {Swank}, {Smith}, {Hurley}, \&
  {Thompson}}]{2001ApJ...552..748W}
{Woods}, P.~M., {Kouveliotou}, C., {G{\"o}{\v g}{\"u}{\c s}}, E., {et~al.}
  2001, \apj, 552, 748

\bibitem[{{Woods} \& {Thompson}(2006)}]{2006csxs.book..547W}
{Woods}, P.~M., \& {Thompson}, C. 2006, {Soft gamma repeaters and anomalous
  X-ray pulsars: magnetar candidates}, ed. W.~H.~G. {Lewin} \& M.~{van der
  Klis}, 547--586

\bibitem[{{Yakovlev} \& {Pethick}(2004)}]{2004ARA&A..42..169Y}
{Yakovlev}, D.~G., \& {Pethick}, C.~J. 2004, \araa, 42, 169

\end{thebibliography}

\clearpage

\end{document}